\documentstyle{article}

\begin{document}
\author{L. Herrera\thanks{%
On leave from Departamento de F\'\i sica, Facultad de Ciencias, Universidad
Central de Venezuela, Caracas, Venezuela and Centro de Astrof\'\i sica
Te\'orica, M\'erida, Venezuela.} \\
\'Area de F\'\i sica Te\'orica\\
Facultad de Ciencias\\
Universidad de Salamanca\\
37008, Salamanca, Espa\~na.\\
and \and J. Mart\'\i nez \\
Grupo de F\'\i sica Estad\'\i stica\\
Departamento de F\'\i sica\\
Universidad Aut\'onoma de Barcelona\\
08193 Bellaterra, Barcelona, Espa\~na.}
\title{Dissipative Collapse Through the Critical Point}

\date{}

\maketitle

\begin{abstract}
A relativistic model of a heat conducting collapsing star, which includes
thermal pre-relaxation processes, is presented. Particular attention is paid
to the influence of a given parameter defined in terms of thermodynamic
variables, on the outcome of evolution. Evaluation of the system when
passing through a critical value of the aforesaid parameter, does not yield
evidence of anomalous behaviour.
\end{abstract}

{\bf{keywords:}Self-gravitating systems, Relativistic stars: stability, Late stages
of stellar evolution, Relativistic fluid dynamics

\newpage

\section{Introduction}

The behaviour of dissipative systems at the very moment when they depart
from hydrostatic equilibrium has been recently studied \cite
{Herrera97,HeMa97a,visco}.

In \cite{Herrera97} it appears that a parameter $\alpha $ formed by a
specific combination of thermal conductivity coefficient $\kappa$, 
relaxation time $\tau$, temperature $T$, proper energy density $\rho$
and pressure $p$, 
\[
\alpha =\frac{\kappa T}{\tau \left( \rho +p\right) }, 
\]
may critically affect the evolution of the object. Specifically, it was
shown that in the equation of motion of any fluid element, the inertial mass
density term vanishes for $\alpha =1$ (critical point) and is negative
beyond that value.

In some cases (pure shear or bulk viscosity) \cite{visco}, the critical
point is well beyond the point where the causality requirements are violated
and therefore forbidden.

In others (pure thermal conduction) \cite{Herrera97,HeMa97a}, later
requirements are violated slightly below the critical point.

In the general case ( heat conduction plus viscosity) it appears that
causality may break down beyond the critical point \cite{visco}.

However, it should be stressed that the critical point (as well as causality
conditions) is obtained in the context of a linear perturbative scheme,
where the system is evaluated immediatly after leaving the equilibrium and
therefore time derivative of radial velocity as well as dissipative
variables are considered small quantities, such that only linear expressions
of them are kept. On the other hand the vanishing, at the critical point, of
the inertial mass density term, leading to accelerated fluid elements in the
absence of total radial forces, might suggest that linear approximation is
not longer reliable there \cite{HeMa97a,visco}.

The question arises then, whether a physical system may reach the critical
point without exhibiting a clear unphysical behavior or if, on the contrary,
for any physical system reaching that point the march of physical variables
becomes physically unacceptable as suggested by the linear approximation. In
other words we want to see how the physical meaning of the above mentioned
parameter $\alpha $, as implied by the linear approximation, carries over to
nonlinear regimes.

In order to elucidate this question we shall consider here an exact
numerical model of an evolving dissipative star (without viscosity). The
system is forced to evolve through the critical point, and fundamental
variables are monitored to detect any anomalous behaviour. Modeling is
performed by means of the HJR method \cite{HeJiRu80}, a brief resume of
which is given in next section .

The heat conduction equation is given in section 3, and the model is
described in section 4. Finally, a brief analysis of our results is given in
the last section.

\section{The HJR method}

We shall consider a non-static spherically symmetric distribution of matter
which consists of fluid, which may be locally anisotropic, and heat flow
(radiation). Assuming Bondi coordinates \cite{BoBuMe62,Bondi64}, the metric
takes the form 
\begin{equation}
ds^2=e^{2\beta }\left[ \frac Vrdu^2+2dudr\right] -r^2\left( d\theta
^2+sin^2\theta d\phi ^2\right) ,  \label{Bondi}
\end{equation}
where $u~$is a time like coordinate , $r$ is the null coordinate and $\theta 
$ and $\phi $ are the usual angle coordinates. A generalization of the
``mass aspect'' defined by Bondi et al \cite{BoBuMe62} can be introduced by
means of function $\tilde{m}(u,r)$ 
\begin{equation}
V=e^{2\beta }(r-2\tilde{m}(u,r)),
\end{equation}
where $\beta $ and $V$ are functions of $u$ and $r$. Inside the fluid
distribution, the stress-energy tensor can be written as -see \cite
{Martinez96} for details- 
\begin{equation}
T_{\mu \nu }=(\rho +P_{\perp })U_\mu U_\nu -P_{\perp }g_{\mu \nu
}+(P_r-P_{\perp })\chi _\mu \chi _\nu +2q_{(\mu }U_{\nu )},
\end{equation}
where $\rho $, $P_r$, $P_{\perp }$ are the energy density, radial pressure
and tangential pressure respectively as measured by a Minkowskian observer
in the Lagrangean frame, and $\chi _\mu =-q_\mu /q$, being $q=\sqrt{-q^\mu
q_\mu }$ the heat flow. Using a Lorentz transformation in the radial
direction and the coordinate transformation between Bondi coordinates and
local Minkowskian coordinates 
\begin{equation}
dt=e^\beta (\sqrt{\frac Vr}du+\sqrt{\frac rV}dr),  \label{ccoor}
\end{equation}
\begin{equation}
dx=e^\beta \sqrt{\frac rV}dr,  \label{xccor}
\end{equation}
\begin{equation}
dy=rd\theta ,  \label{ycoor}
\end{equation}
\begin{equation}
dz=rsin\theta d\phi ,  \label{zcoor}
\end{equation}
it is possible to express the components of the stress-energy tensor in
Bondi coordinates, in terms of variables measured in the Minkowskian
Lagrangean frame. In Bondi coordinates the four-velocity $U_\mu $, and the
heat flux vector $q^\mu $ are given by 
\begin{equation}
U_\mu =e^\beta \left( \sqrt{\frac Vr}\frac 1{(1-\omega ^2)^{1/2}},\sqrt{%
\frac rV}\left( \frac{1-\omega }{1+\omega }\right) ^{1/2},0,0\right) ,
\label{umu}
\end{equation}
and 
\begin{equation}
q^\mu =qe^{-\beta }\left( -\sqrt{\frac rV}\left( \frac{1-\omega }{1+\omega }%
\right) ^{1/2},\sqrt{\frac Vr}\frac 1{(1-\omega ^2)^{1/2}},0,0\right) ,
\label{qmu}
\end{equation}
where $\omega $ is the velocity of matter as measured by the Minkowskian
observer defined by (\ref{ccoor})-(\ref{zcoor}).

At the outside of the fluid distribution the space-time is described by the
Vaidya metric \cite{Vaidya51}, which in Bondi coordinates is given by $\beta
=0$ and $V=r-2m(u).$

The Einstein field equations, inside the matter distribution, can be written
as \cite{HeJiEs87}:

\begin{equation}
\frac 1{4\pi r(r-2\tilde{m})}\left( -\tilde{m}_0e^{-2\beta }+(1-2\tilde{m}/r)%
\tilde{m}_1\right) =\frac 1{1-\omega ^2}(\rho +2\omega q+P_r\omega ^2),
\label{ecu00}
\end{equation}
\begin{equation}
\frac{\tilde{m}_1}{4\pi r^2}=\tilde{\rho},  \label{ecu01}
\end{equation}
\begin{equation}
\beta _1\frac{r-2\tilde{m}}{2\pi r^2}=\tilde{\rho}+\tilde{P},  \label{ecu11}
\end{equation}
\begin{equation}
-\frac{\beta _{01}e^{-2\beta }}{4\pi }+\frac 1{8\pi }(1-2\frac{\tilde{m}}r%
)(2\beta _{11}+4\beta _1^2-\frac{\beta _1}r)+\frac{3\beta _1(1-2\tilde{m}_1)-%
\tilde{m}_{11}}{8\pi r}=P_{\perp },  \label{ecu22}
\end{equation}
where subscripts 0 and 1 denote partial derivative with respect to $u$ and $%
r $ respectively, and the effective energy density 
\begin{equation}
\tilde{\rho}=\frac 1{1+\omega }(\rho -q(1-\omega )-P_r\omega ),
\label{rotilde}
\end{equation}
and the effective pressure 
\begin{equation}
\tilde{P}=\frac 1{1+\omega }(-\omega \rho -q(1-\omega )+P_r)  \label{ptilde}
\end{equation}
are two auxiliary functions introduced in the HJR formalism \cite
{HeJiRu80,HeJiEs87,CoHeEsWi82} whose physical meaning becomes clear in the
static case, in which they reduce to the energy density and radial pressure
respectively.

Matching the Vaidya metric to the Bondi metric at the surface ($r=a$) of the
fluid distribution implies -- see \cite{HeJi83} for details -- 
\begin{equation}
\tilde{P}_a=-\omega _a\tilde{\rho}_a,  \label{jc5}
\end{equation}
which is equivalent to the well-known condition 
\begin{equation}
q_a=P_{ra},  \label{jc6}
\end{equation}
for radiative spheres \cite{Santos85}, (subscript a indicates that the
quantity is evaluated at the boundary surface).

The HJR method is based on a system of three ordinary differential equations
for quantities evaluated on the boundary surface (surface equations), which
will allow us to find the evolution of the physical quantities.

The dimensionless variables 
\begin{equation}
A\equiv \frac a{\widetilde{m}(0)}\;\;\;\;M\equiv \frac{\widetilde{m}}{%
\widetilde{m}(0)}\;\;\;\;u\equiv \frac u{\widetilde{m}(0)}\;\;\;\;F\equiv 1-%
\frac{2M}A\;\;\;\;\Omega \equiv \frac 1{1-\omega _a},  \label{adime}
\end{equation}
are defined to derive the surface equations, where $\widetilde{m}(0)$ is the
initial total mass of the system. Then, using (\ref{adime}) and boundary
conditions one obtains the first surface equation 
\begin{equation}
\dot{A}=F(\Omega -1),  \label{se1}
\end{equation}
which gives the evolution of the radius of the star, (where dot denotes
derivative with respect to the dimensionless $u$).

The second surface equation 
\begin{equation}
\dot{F}=\frac{2L+F(1-F)(\Omega -1)}A,  \label{se2}
\end{equation}
gives the evolution of the redshift at the surface -see \cite
{HeJiEs87,HeJiRu80,CoHeEsWi82} for details. The luminosity, $L$, as measured
by an observer at rest at infinity reads 
\begin{equation}
L=-\dot{M}=\frac E{(1+z_a)^2}=EF=\hat{E}(2\Omega -1)F=4\pi A^2q_a\left(
2\Omega -1\right) F,  \label{luminosity}
\end{equation}
where $z_a$ refers to the boundary gravitational redshift, $\hat{E}$ is the
luminosity as seen by a comoving observer, and $E$ is the luminosity
measured by a non comoving observer located on the surface.

The third surface equation is model dependent. For anisotropic fluids the
relationship $(T_{r;\mu }^\mu )_a=0$ can be written as 
\[
\frac{\dot F}F+\frac{\dot \Omega }\Omega -\frac{\stackrel{.}{\tilde \rho }_a%
}{\tilde \rho _a}+F\Omega ^2\frac{\tilde R_{\perp _a}}{\tilde \rho _a}-\frac 
2AF\Omega \frac{P_{ra}}{\tilde \rho _a}= 
\]
\begin{equation}  \label{se3}
(1-\Omega )\left[ 4\pi A\tilde \rho _a\frac{3\Omega -1}\Omega -\frac{3+F}{2A}%
+F\Omega \frac{\tilde \rho _{1_a}}{\tilde \rho _a}+\frac{2F\Omega }{A\tilde 
\rho _a}(P_{\perp }-P_r)_a\right] ,
\end{equation}
being 
\begin{equation}
\tilde R_{\perp _a}=\tilde P_{1_a}+\left( \frac{\tilde P+\tilde \rho }{1-2%
\tilde m/r}\right) _a\left( 4\pi r\tilde P+\frac{\tilde m}{r^2}\right)
_a-\left( \frac 2r(P_{\perp }-P_r)\right) _a.
\end{equation}
Expression (\ref{se3}) generalizes the Tolman-Oppenheimer-Volkov equation to
the non-static radiative anisotropic case.

The HJR method \cite{HeJiRu80} allows us to find non static solutions of the
Einstein equations from static ones. The algorithm, extended for anisotropic
fluids, can be found in \cite{CoHeEsWi82}. Nevertheless, for our purpose
here, it is only necessary to find the evolution of $A$, $F$, $\Omega $, and 
$L$. Thus, the algorithm can be resumed as follows

\begin{enumerate}
\item  Take a static but otherwise arbitrary interior solution of the
Einstein equations for a spherically symmetric fluid distribution (''seed''
solution) 
\begin{equation}
P_{st}=P(r),\ \ \ \ \ \rho _{st}=\rho (r).
\end{equation}

\item  Assume that the $r$-dependence of the effective quantities is the
same as that of the energy density and radial pressure of the ''seed''
solution. Nevertheless, note that junction conditions in terms of effective
variables, read as (\ref{jc5}). This condition allows us to find out the
relation between the $u$-dependence of $\tilde{\rho}\equiv \tilde{\rho}(u,r)$
and $\tilde{P}\equiv \tilde{P}(u,r)$.

\item  If we have an expression for $\tilde{\rho}(u,r=a)$ and $\tilde{P}%
(u,r=a),$ it is possible to solve the system of surface equations (\ref{se1}%
), (\ref{se2}), and (\ref{se3}) for a given luminosity.
\end{enumerate}

\section{Heat conduction equation}

As mentioned in the Introduction, the value of parameter $\alpha $, defined
as 
\begin{equation}
\alpha =\frac{\kappa T}{\tau (\rho +p)}  \label{alfa0}
\end{equation}
appears to be important in the evolution of radiating stars. In fact, as
it is shown in \cite{Herrera97,HeMa97a}, if the system reaches the critical
point ($\alpha =1$), then the inertial mass term vanishes. Furthermore it
can be shown that conditions ensuring stability and causality \cite
{His83,Maar.96} are violated at the critical point \cite{HeMa97a}. This
strange result may have, at least, two different interpretations. On one
hand, it may happen that, because of the fact that the inertial mass term
vanishes at the critical point, a linear perturbative scheme fails at the
critical point, invalidating thereby any restriction obtained from linear
aproximation (e.g. causality and stability conditions mentioned above).
Alternatively, it may happen that the system is actually prevented from
reaching the critical point, in which case it should exhibit some abnormal
behaviour when approaching the critical point. The correct interpretation
can be found by solving the Einstein field equations, together with the
transport equation, without applying linear perturbation theory.

In order to study the behaviour of the system at the critical point, it is
convenient to adopt a physical framework in which condition $\alpha =1$
could be overtaken for reasonable values of physical quantities. A good
candidate seems to be a collapsing neutron star in which the equilibrium is
reached by means of a huge emission of neutrinos. The large temperature
reached during the process of collapse would explain values of $\alpha $
greater than $1.$

We assume the evolution of the heat flow to be governed by the
Maxwell-Cattaneo transport equation, 
\begin{equation}
\tau h_\nu ^\mu \stackrel{.}{q}^\nu +q^\mu =\kappa h^{\mu \nu }\left(
T_{,\nu }-T\stackrel{.}{U}_\nu \right) ,  \label{mc}
\end{equation}
where 
\begin{equation}
\stackrel{.}{U}_\nu =U^\alpha U_{\nu ;\alpha },  \label{upun}
\end{equation}
\begin{equation}
\stackrel{.}{q}_\nu =U^\alpha q_{\nu ;\alpha },  \label{qpun}
\end{equation}
and $\kappa $ and $\tau $ denote the thermal conductivity coefficient and
relaxation time respectively. Evaluating (\ref{mc}) in the surface, the
transport equation can be written as 
\[
\tau \stackrel{.}{q}_a+q_a\sqrt{F(2\Omega -1)}= 
\]
\begin{equation}
\kappa _a\left[ \stackrel{.}{T}_a-T_{1a}F(2\Omega -1)-T_a\left( \frac{\Omega
(1-F)}{2A}+\frac L{AF}+\frac{\stackrel{.}{\Omega }}{2\Omega -1}\right)
\right] .  \label{mca}
\end{equation}

The thermal conductivity coefficient for a mixture of matter and radiation
is given by \cite{Weinberg71} 
\begin{equation}
\kappa =\frac 43bT^3\tau _{col},  \label{kappa}
\end{equation}
where $b=7N_\nu a/8$ for neutrinos, being $a$ the radiation constant, $N_\nu 
$ the number of neutrino flavors and $\tau _{col}$ the matter-neutrinos time
collision.

On the other hand, the luminosity perceived by a comoving observer located
on the surface can be connected with the effective temperature, $T_{eff}$,
as 
\begin{equation}
\widehat{E}=\frac L{F(2\Omega -1)}=4\pi A^2q_a=\left[ 4\pi r^2\sigma
T_{eff}^4\right] _{r=a},  \label{E}
\end{equation}
where $\sigma =a/4$ is the Steffan-Boltzman radiation constant. The concept
of effective temperature has been widely used in theory of stellar
atmospheres \cite[p.586]{ShTe83}, \cite[p.70]{KiWe94}, and \cite[p.295]
{HaKa94}. The effective temperature can be connected to the material
temperature by means of 
\begin{equation}
T_a^4=\frac 12\left[ T_{eff}^4\right] _{r=a}.  \label{ta}
\end{equation}
Thus, from (\ref{E}) and (\ref{ta}), the dimensionless heat flow at the
surface is 
\begin{equation}
q_a=\frac{2\zeta \xi ^2}{2\Omega -1}T_a^4,  \label{qa}
\end{equation}
where $\zeta =\sigma M_{\odot }^2\simeq 3.4097\times 10^{-54}$ K$^{-4}$, and 
$\xi =M_o/M_{\odot }.$ In the HJR formalism the initial mass is normalized
to unity. Thus, every term in (\ref{mca}) is dimensionless and $\kappa $
must be expressed as 
\begin{equation}
\kappa =\frac{14}3\zeta T^3\xi ^2N_\nu \tau _{col}.  \label{newk}
\end{equation}

On the other hand, evaluating (\ref{alfa0}) at the surface, and by means of
the definition of (\ref{rotilde}), (\ref{jc6}) and (\ref{newk}), it takes
the form 
\begin{equation}
\alpha =\frac{14y^4\xi ^2N_\nu \Omega (2\Omega -1)}{3\left( \widetilde{\rho }%
_a(2\Omega -1)^2+4y^4\xi ^2\Omega \right) },  \label{alfa}
\end{equation}
where we have assumed $\tau \sim \tau _{col}$, and the parameter $y$ has
been defined as 
\begin{equation}
y^4=\zeta T_a^4.  \label{y4}
\end{equation}

Now, it is possible to write $y$ in terms of $\alpha $ by means of (\ref
{alfa}) 
\begin{equation}
y^4=\frac{3\widetilde{\rho }_a(2\Omega -1)^2\alpha }{2\xi ^2\Omega \left(
7N_\nu (2\Omega -1)-6\alpha \right) },  \label{y4a}
\end{equation}
and by substitution of this expression and (\ref{y4}), into (\ref{qa}) 
\begin{equation}
q_a=\frac{3\widetilde{\rho }_a(2\Omega -1)\alpha }{\Omega \left( 7N_\nu
(2\Omega -1)-6\alpha \right) }.  \label{qaa}
\end{equation}
After some elementary algebra, expression (\ref{mca}) can be written in
terms of $\alpha $ instead of $y$ and $q_a$%
\[
\frac{y_1}yF(2\Omega -1)=\frac{\stackrel{.}{\Omega }}{2\Omega -1}\left( 
\frac 6{\Phi +6\alpha }-1\right) +\frac{\Omega (F-1)}{2A}-\frac L{AF}-\frac{3%
\sqrt{F}}{7\tau _{col}N_\nu \sqrt{2\Omega -1}}+ 
\]
\begin{equation}
\left( \frac{\stackrel{.}{\widetilde{\rho }}_a}{\widetilde{\rho }_a}+\frac{%
\stackrel{.}{\alpha }}\alpha \left[ 1+\frac{6\alpha }\Phi \right] +\frac{%
\stackrel{.}{\Omega }\left( \Phi -12\alpha \Omega \right) }{\Omega \Phi
(2\Omega -1)}\right) \left[ \frac 14-\frac 3{\Phi +6\alpha }\right] ,
\label{gradt}
\end{equation}
where $\Phi =7N_\nu (2\Omega -1)-6\alpha .$

Our purpose is to discern what anomalous effects (if any) can take place
when the system overtakes the critical point. Thus, it seems reasonable to
impose a profile for $\alpha $, and study the evolution of the system. The
Maxwell-Cattaneo transport equation (\ref{gradt}) allows us to find the
temperature gradient in the surface for a given $\alpha (u).$ At first
glance, expression (\ref{gradt}) does not seems to present anomalous
behaviour in the critical point. Nevertheless, it is necessary to study the
complete evolution of the system to confirm this suspicion. The method that
we shall use can be summarized as follows

\begin{enumerate}
\item  Impose a profile for $\alpha $. According to (\ref{alfa0}), $\alpha $
must vanish initially if the system departs from equilibrium ($T_a(u=0)\sim
0 $)

\item  For a given $\alpha $, it is possible to find the luminosity $L$ by
means of (\ref{E}) and (\ref{qaa}).

\item  Follow the HJR method, outlined above, to solve the system of surface
equations.

\item  Once $A$, $F$, and $\Omega $ are found it is possible to find the
temperature gradient by means of (\ref{gradt}). Assuming that cooling by
absortion and emission of neutrinos is the responsible to drive the star to
a new equilibrium state, the mean collision time $\tau _{col}$ can be
roughly expressed as \cite{Martinez96} 
\begin{equation}
\tau _{col}\sim {\cal {A}}\frac{M_{\odot }\xi \zeta ^{3/8}}{\rho \sqrt{{\cal 
{Y}}y^3}},  \label{tcol}
\end{equation}
where ${\cal {A}}=10^9$ K$^{3/2}$m$^{-1},$ and ${\cal {Y}}$ stands for the
electron fraction.
\end{enumerate}

\section{The model}

In order to study the evolution of the system beyond the critical point we
adopt the Gokhroo \& Mehra-type solution \cite{GoMe94} as the "seed"
solution. This solution corresponds to an anisotropic fluid with
inhomogeneous energy density, and it can take account of the large density
and pressure gradients close to surface \cite{Martinez96}.

It can be shown \cite{Martinez96} that the effective energy density and
effective pressure are given for this model by 
\begin{equation}
\widetilde{\rho }=\rho _c\frac{K(u)}{K_o}\left( 1-K(u)\frac{r^2}{A^2}\right)
,  \label{tilderho}
\end{equation}
and 
\begin{equation}
\widetilde{P}=\lambda \rho _c\frac{K(u)}{K_o}\left( 1-2\frac{\widetilde{m}}r%
\right) \left( 1-G(u)\frac{r^2}{A^2}\right) ^n,  \label{tildepres}
\end{equation}
where $n\geq 1$, and the central energy density in the static case $\rho _c$
is given by 
\begin{equation}
\rho _c=\frac{15}{4\pi A_o(5-3K_o)}  \label{roc}
\end{equation}

If we define 
\begin{equation}
\gamma =\frac{8\pi \rho _c}3,  \label{alpha}
\end{equation}
then functions $K(u)$, and $G(u)$ are 
\begin{equation}
K(u)=\left\{ 
\begin{array}{cc}
\frac 56\left[ 1+\sqrt{1-\left( \frac{12K_o}{5\gamma }\right) \left( \frac{%
1-F}{A^2}\right) }\right] & \mbox{if }K_o>\frac 56 \\ 
&  \\ 
\frac 56\left[ 1-\sqrt{1-\left( \frac{12K_o}{5\gamma }\right) \left( \frac{%
1-F}{A^2}\right) }\right] & \mbox{if }K_o<\frac 56
\end{array}
\right. ,  \label{K}
\end{equation}
and 
\begin{equation}
G(u)=1-\left[ \frac{\left( 1-\Omega \right) \left( 1-K\right) }{F\Omega
\lambda }\right] ^{1/n}.  \label{G}
\end{equation}

Thus, the system of surface equations for this model is 
\begin{equation}
\stackrel{.}{A}=F(\Omega -1),  \label{se1go}
\end{equation}
\begin{equation}
\stackrel{.}{F}=\frac 1A\left[ 2L+F(1-F)(\Omega -1)\right] ,  \label{se2go}
\end{equation}
\begin{equation}
\stackrel{.}{\Omega }=-\frac{\stackrel{.}{F}}F\Omega +\frac{\stackrel{.}{K}}K%
\frac{(1-2K)}{(1-K)}\Omega +\frac{4K_oL\Omega ^2}{3\gamma KA^3(2\Omega
-1)(1-K)}+\Omega (1-\Omega )\Lambda ,  \label{se3go}
\end{equation}
where 
\begin{equation}
\Lambda =\frac{3\gamma K}{2K_o}A(1-K)\left( \frac{3\Omega -1}\Omega \right) -%
\frac{3+F}{2A}+\frac{2F\Omega }{A(1-K)}(\Psi -K),
\end{equation}
\[
\Psi =\frac 3{10K_o}\lambda \gamma A^2K^2\left( 1-G\right) ^n 
\]
\begin{equation}
+\frac{A^2}{2F}\left[ \frac{3\gamma K}{2K_o}\lambda ^2F^2\left( 1-G\right)
^{2n}-\frac{2n\lambda G}{A^2}F^2\left( 1-G\right) ^{n-1}+\frac \gamma 2%
\left( 1-\frac{3K}5\right) \frac{K\left( 1-K\right) }{K_o}\right] .
\end{equation}

The mean collision time (\ref{tcol}) can be rewritten in terms of $\Omega $, 
$\alpha $ and $\widetilde{\rho }$ as 
\begin{equation}
\tau _{col}=\frac{{\cal {A}}M_{\odot }}{\sqrt{{\cal {Y}}}\left( \Phi
+3\alpha \right) }\left( \frac{\Omega \Phi }{\widetilde{\rho }_a}\right)
^{11/8}\left( \frac \xi {2\Omega -1}\right) ^{7/4}\left[ \frac{2\zeta }{%
3\alpha }\right] ^{3/8},  \label{tcol2}
\end{equation}
whereas the luminosity can be found from (\ref{E}) and (\ref{qaa}) 
\begin{equation}
L=\frac{12\pi A^2\widetilde{\rho }_aF(2\Omega -1)^2\alpha }{\Omega \Phi }.
\label{newlumino}
\end{equation}
Thus, the temperature gradient in the surface can be found by means of (\ref
{tcol2}), (\ref{newlumino}) and (\ref{gradt}) for a given $N_\nu .$

The system of equations (\ref{se1go}-\ref{se3go}) and (\ref{gradt}) has been
solved for two set of initial values:

\begin{enumerate}
\item  Model with initial values $A_o=A(u=0)=6$, $\Omega _o=\Omega (u=0)=1$, 
$n=1$, $K_o=0.999$, $\lambda =1/3$, $N_\nu =3$ and ${\cal {Y}}=0.2$. This
initial configuration corresponds to a star initially at rest, with a radius
of $8862$ meters, a central density equal to $1.70\times 10^{15}$ g cm$^{-3}$%
, and a surface density equal to $1.70\times 10^{12}$ g cm$^{-3}$. Its
initial mass is $1M_{\odot }$, and its surface redshift is close to $0.225$.
We assume the center of the star composed by a highly relativistic Fermi
gas. Thus, $\lambda =1/3$. The profile for $\alpha $ has been assumed as a
gaussian centered in $u=u_{peak}$, {\it {i.e.}} 
\begin{equation}
\alpha =\alpha _{\max }\exp \left( -\frac 12\left[ \frac{u-u_{peak}}\Delta
\right] ^2\right) .  \label{prof1}
\end{equation}
We have taken $u_{peak}=100$ ($\sim 0.49$ msec) and $\Delta =13$ ($\sim
0.064 $ msec). For this values the temperature changes considerably along $1$
msec and $\alpha (0)\sim 0$ as is demanded by condition $T(0)\sim 0$. We
have considered three profiles with $\alpha _{\max }=0.9,$ $1$ and $1.1$
(figure 1).

\item  Model initially in slow contraction with $A_o=5$ ($\simeq 7838$
meters), $\Omega _o=0.99926$ ($\left| \omega \right| \ll 1$), $\xi =n=1$, $%
K_o=0.999$, $\lambda =1/3$, $N_\nu =3$ and ${\cal {Y}}=0.3$. These values
correspond to $\rho _c\simeq 2.94\times 10^{15}$ g cm$^{-3}$ and $\rho
_a\simeq 2.94\times 10^{12}$ g cm$^{-3}$. The profile for $\alpha $ has been
assumed as 
\begin{equation}
\alpha =2-\exp \left( -\frac 12\left[ \frac u\Delta \right] ^2\right) ,
\label{prof2a}
\end{equation}
and 
\begin{equation}
\alpha =1.  \label{prof2b}
\end{equation}
As in the previous case $\Delta =13$ ($\sim 0.064$ msec) (figure 2).
\end{enumerate}

\section{Discussion}

As mentioned before, our purpose here is to clarify the physical
significance (if any) of the critical point in the evolution of a
dissipative self-gravitating system.

In particular we want to find out if at the critical point the system,
described by the full theory (without any approximation), exhibits an
anomalous behaviour, as might suggest a linear approximation approach.

The three profiles of $\alpha$ considered are displayed in figure 1. In one
case the system never reaches the critical point, in other case it is at the
critical point at some moment of its evolution, and in the third case the
system goes beyond the critical point before returning to equilibrium.

Figures (3)-(6) show the evolution of temperature, temperature gradient,
radius and surface velocity for the three different profiles of $\alpha$.

It is apparent from these figures that even though increasing values of $%
\alpha$ are associated with more unstable configurations (in the sense of
faster collapse), nothing strange seems to happens at or beyond the critical
point. Figure (7), showing the evolution of the ratio of the neutrino mean
free path to the radius of the star, indicates that the diffusion
approximation is valid during most part of the emission process.

To reinforce this conclusion we have performed another numerical simulation
with the profiles of $\alpha$ given in figure (2). The evolution of relevant
physical variables displayed in figures (8)-(11), confirms the conclusion
emerging from figures (3)-(6). The validity of diffusion approximation is
corroborated in fig.(12).

To conclude, we may say that the increasing of instability (in the sense
mentioned above) with higher values of $\alpha$, seems to confirm the
decreasing of the inertial mass density factor, predicted in the linear
approximation.

However, nothing dramatic appears at the critical point in the exact
modeling, indicating that the later approximation is not reliable at (or
close to) the critical point, as suggested before \cite{HeMa97a,visco}.

\section*{Acknowledgements}

This work was partially supported by the Spanish Ministry of Education under
Grant No. PB94-0718

\newpage

\section*{Figure captions}

\begin{description}
\item[Figure 1.-]  Profiles of $\alpha $ given by expression (\ref{prof1}).
The values of $\alpha _{\max }$ are $0.9$, $1$ and $1.1.$

\item[Figure 2.-]  Profiles of $\alpha $ for the model initially in slow
contraction (\ref{prof2a}) and (\ref{prof2b}).

\item[Figure 3.-]  Temperature evolution corresponding to three profiles of $%
\alpha $ shown in figure 1.

\item[Figure 4.-]  Temperature gradient for the model initially at rest.

\item[Figure 5.-]  Evolution of the radius for the model initially at rest.

\item[Figure 6.-]  Evolution of the surface velocity for the model initially
at rest.

\item[Figure 7.-]  Mean free path of neutrinos in the first model.

\item[Figure 8.-]  Temperature evolution for the model initially in slow
contraction. The profiles of $\alpha $ corresponding to this model are shown
in figure 2. The curve labeled with $\alpha =1$ corresponds to the profile (%
\ref{prof2b}), whereas the other one corresponds to the profile of $\alpha $
given by expression (\ref{prof2a}).

\item[Figure 9.-]  Temperature gradient for the model initially in slow
contraction.

\item[Figure 10.-]  Evolution of the radius for the model initially in slow
contraction.

\item[Figure 11.-]  Evolution of the surface velocity for the model
initially in slow contraction.

\item[Figure 12.-]  Same as figure 7 for the model initially in slow
contraction.
\end{description}

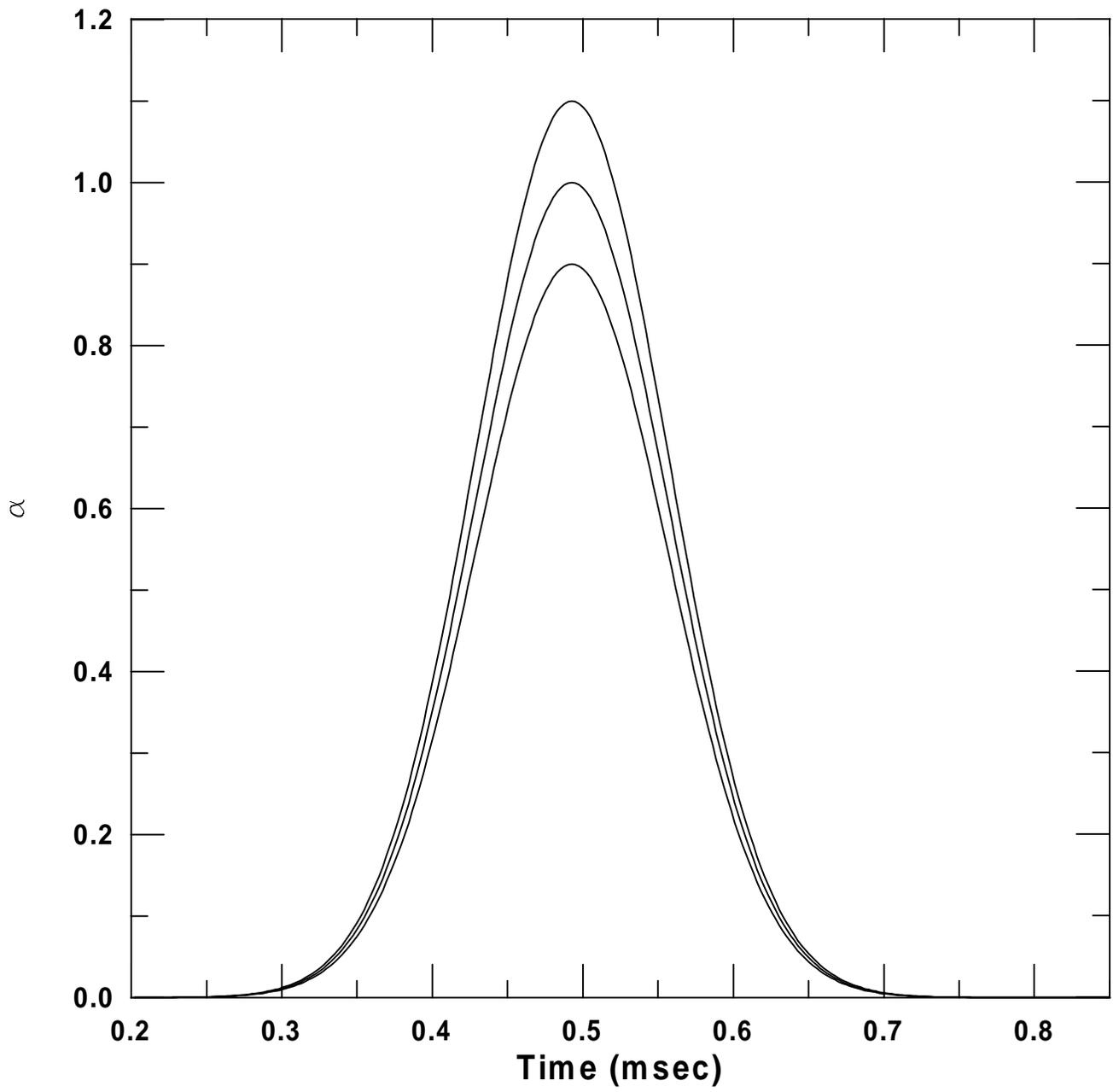
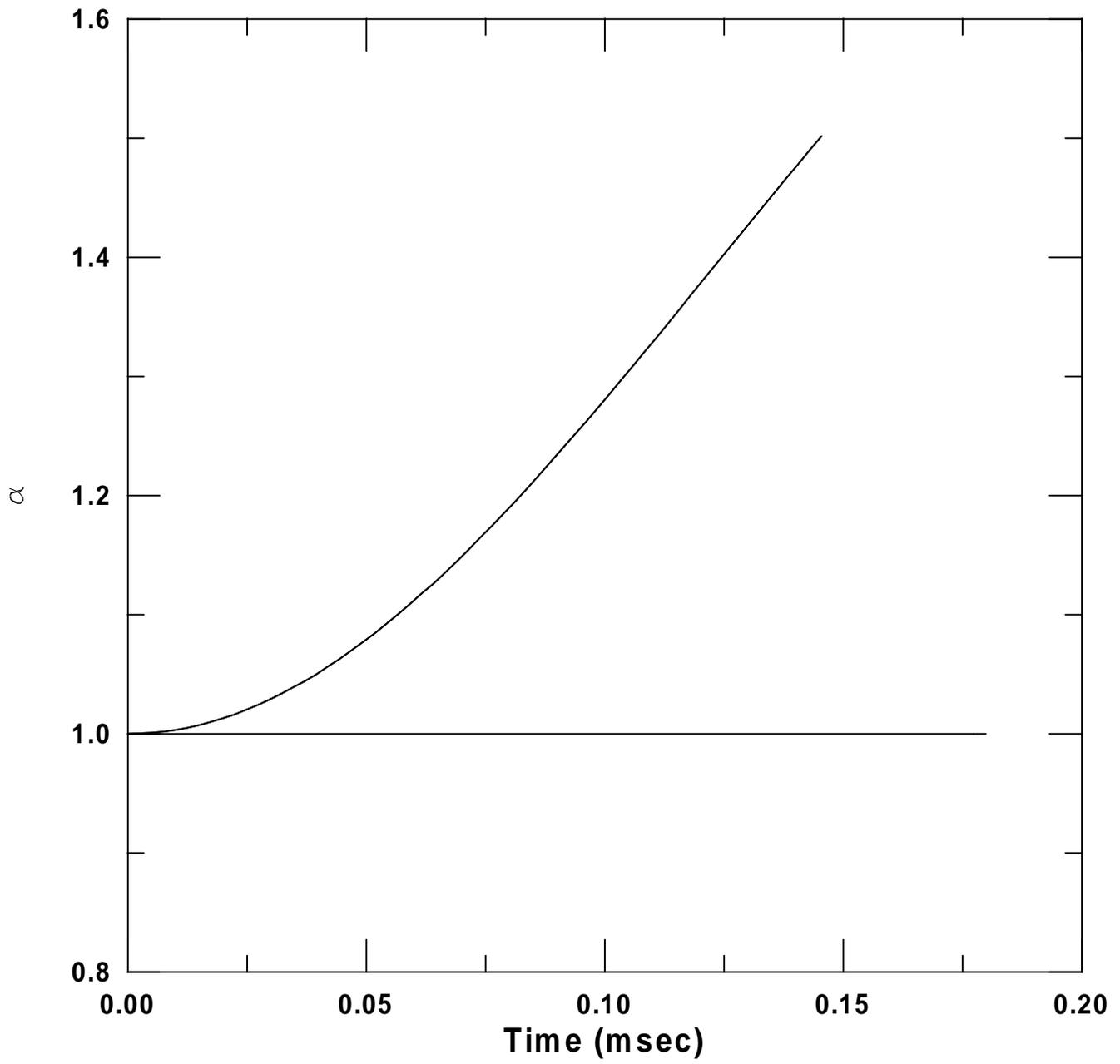
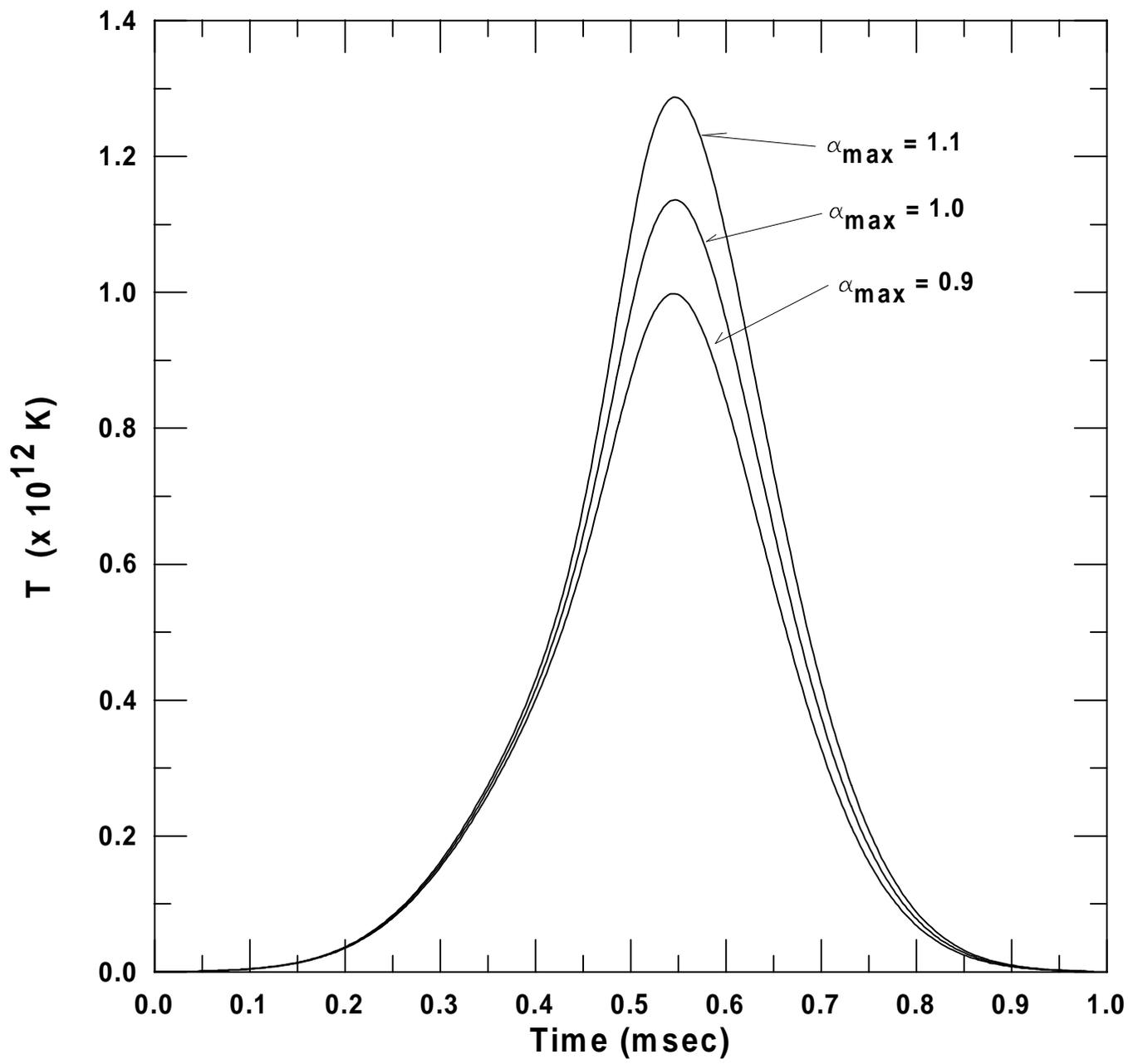
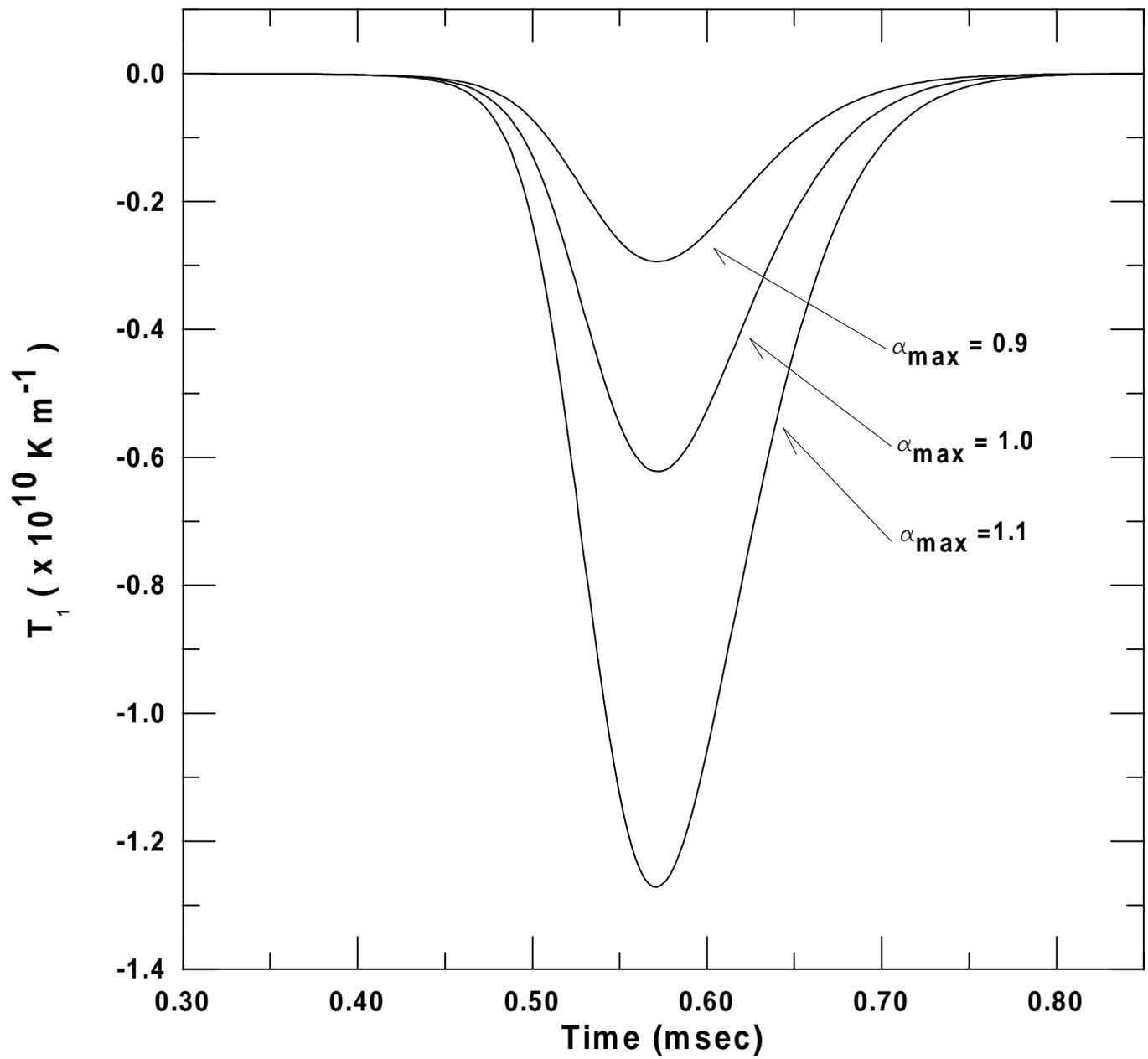
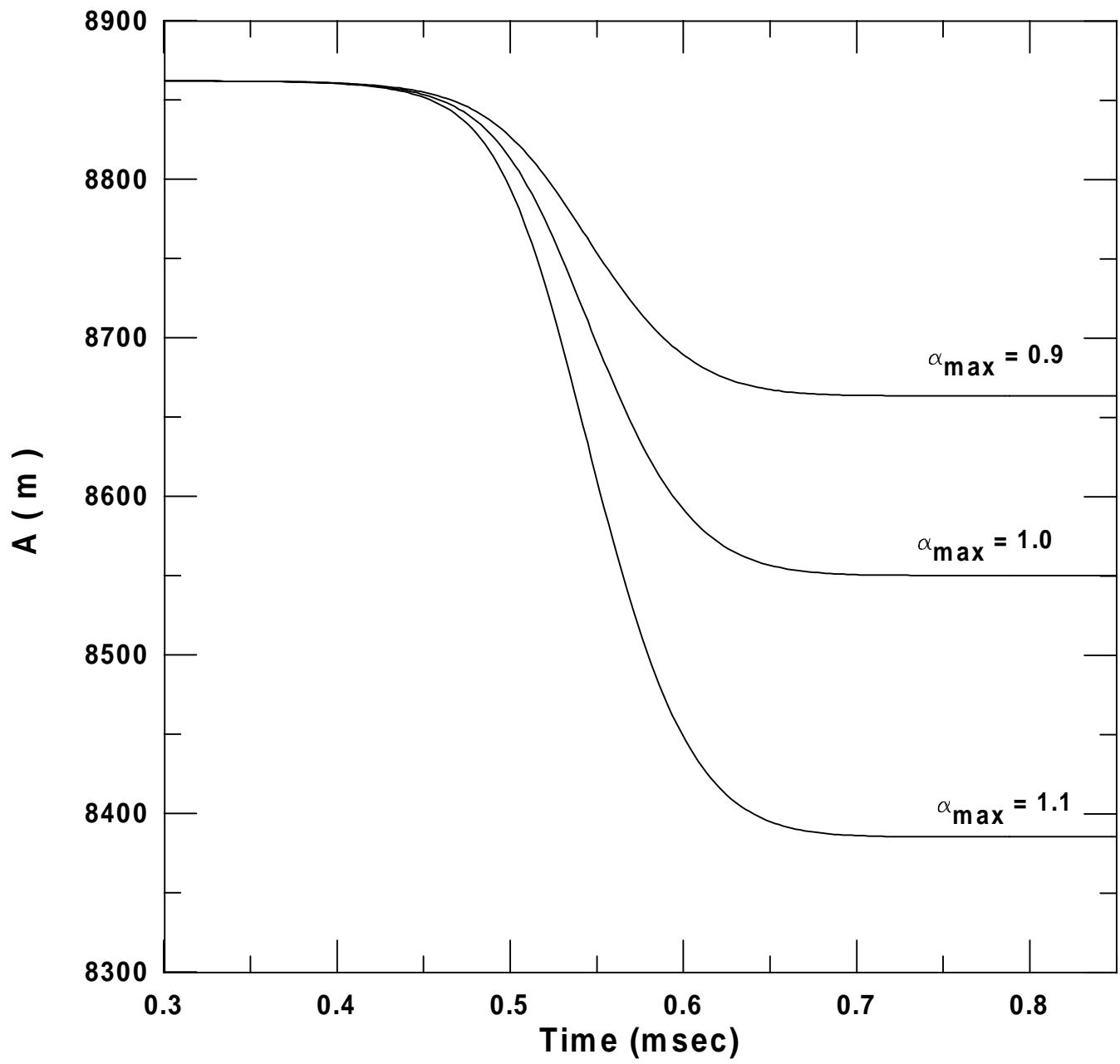
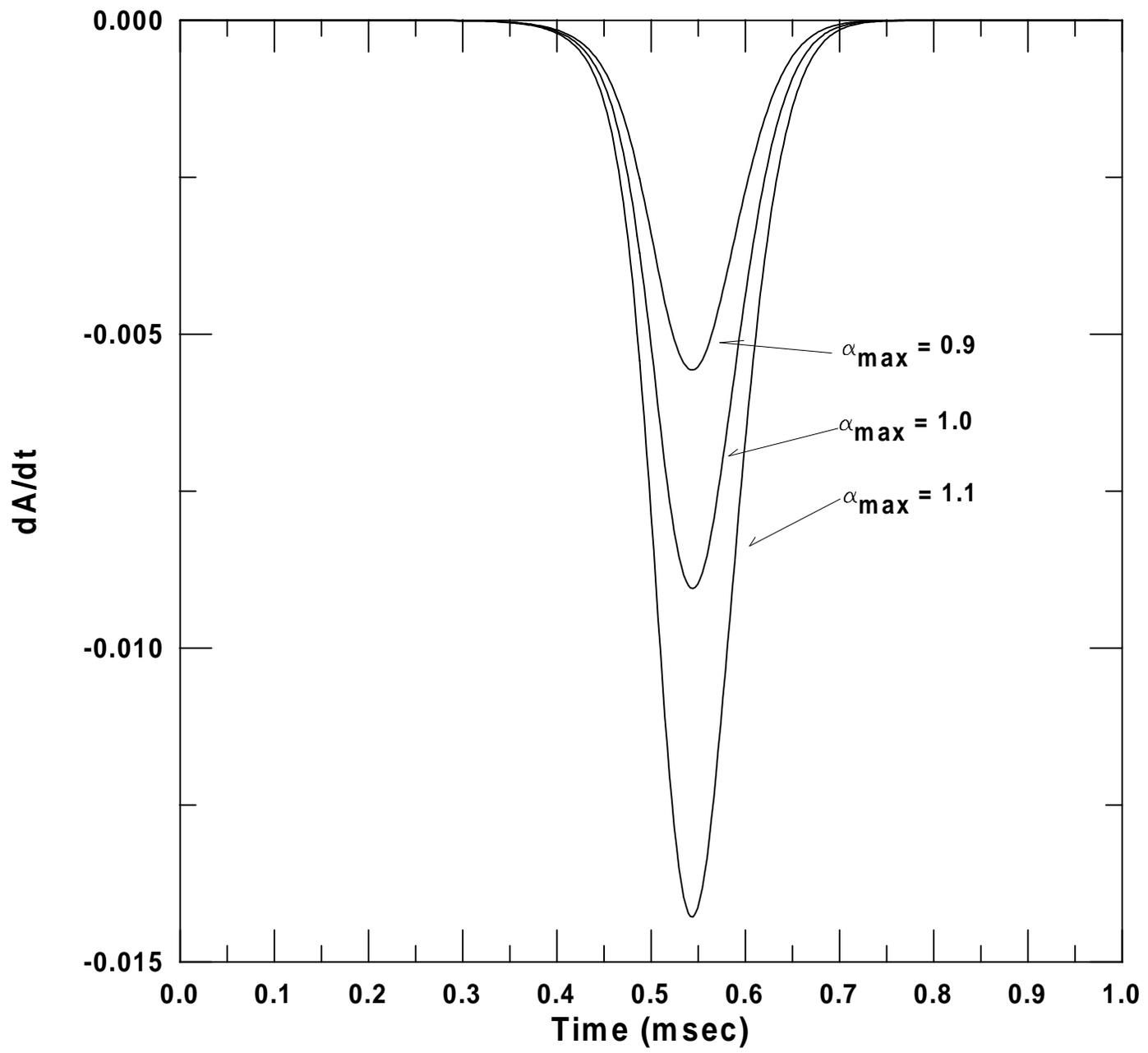
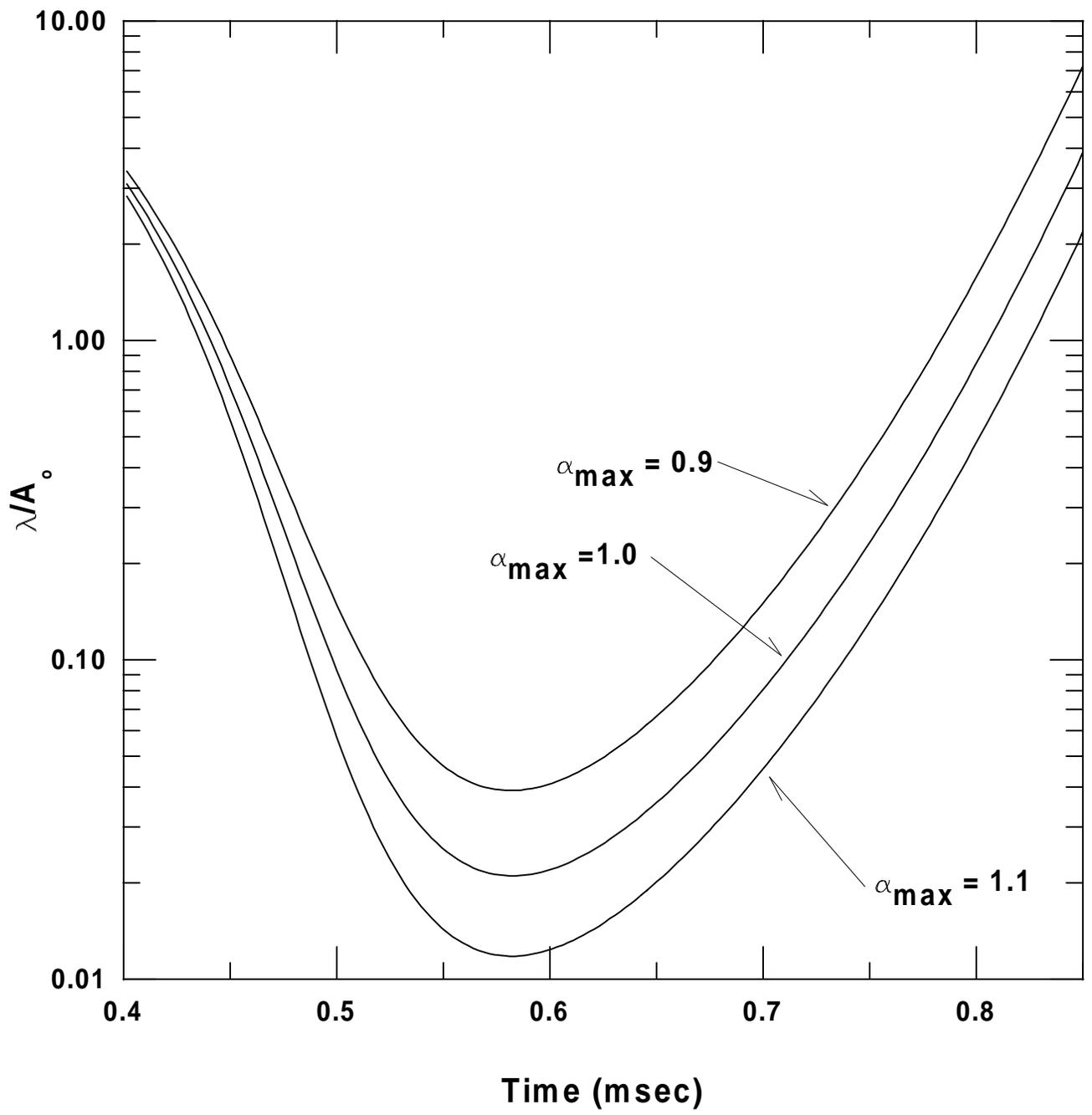
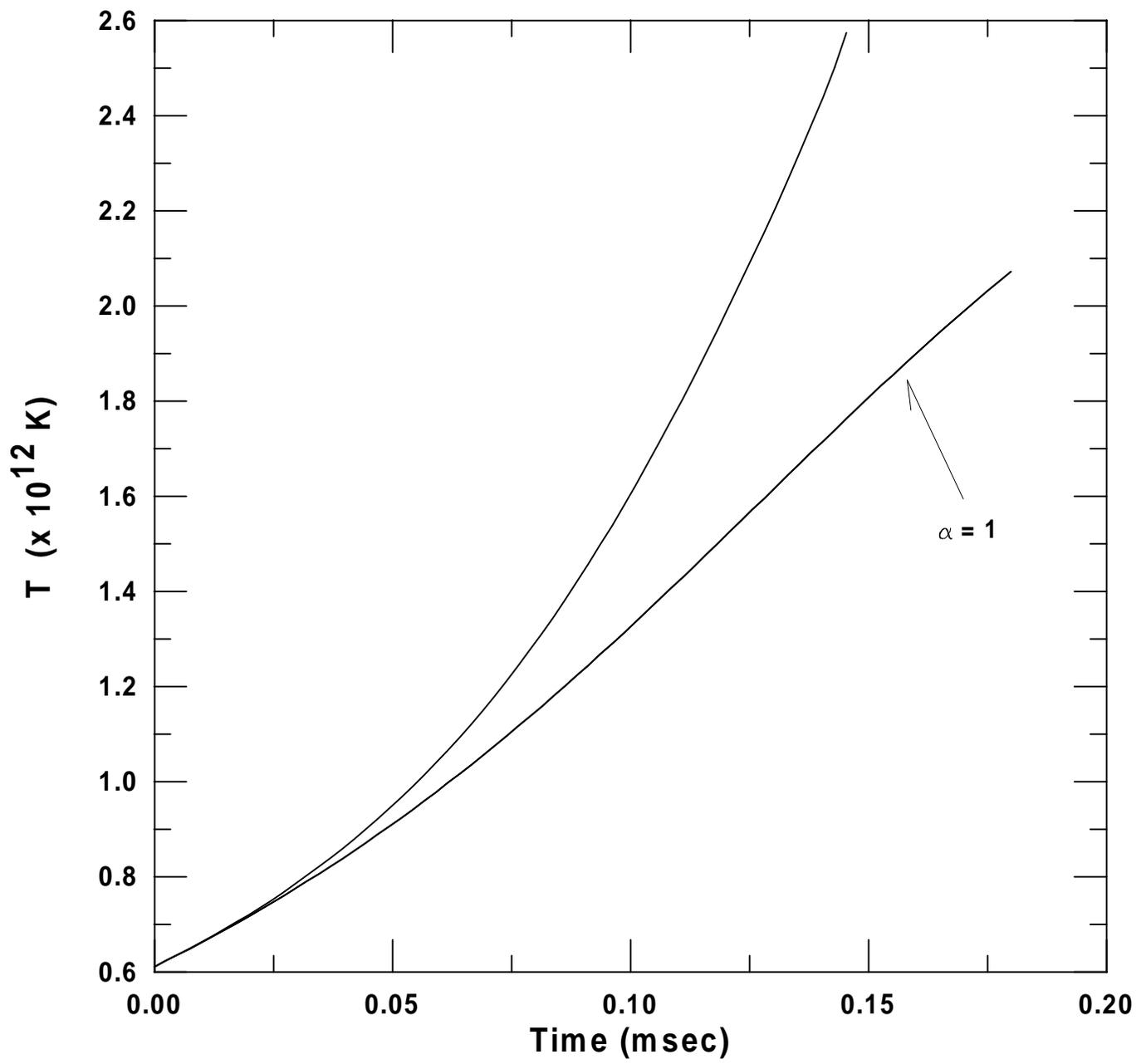
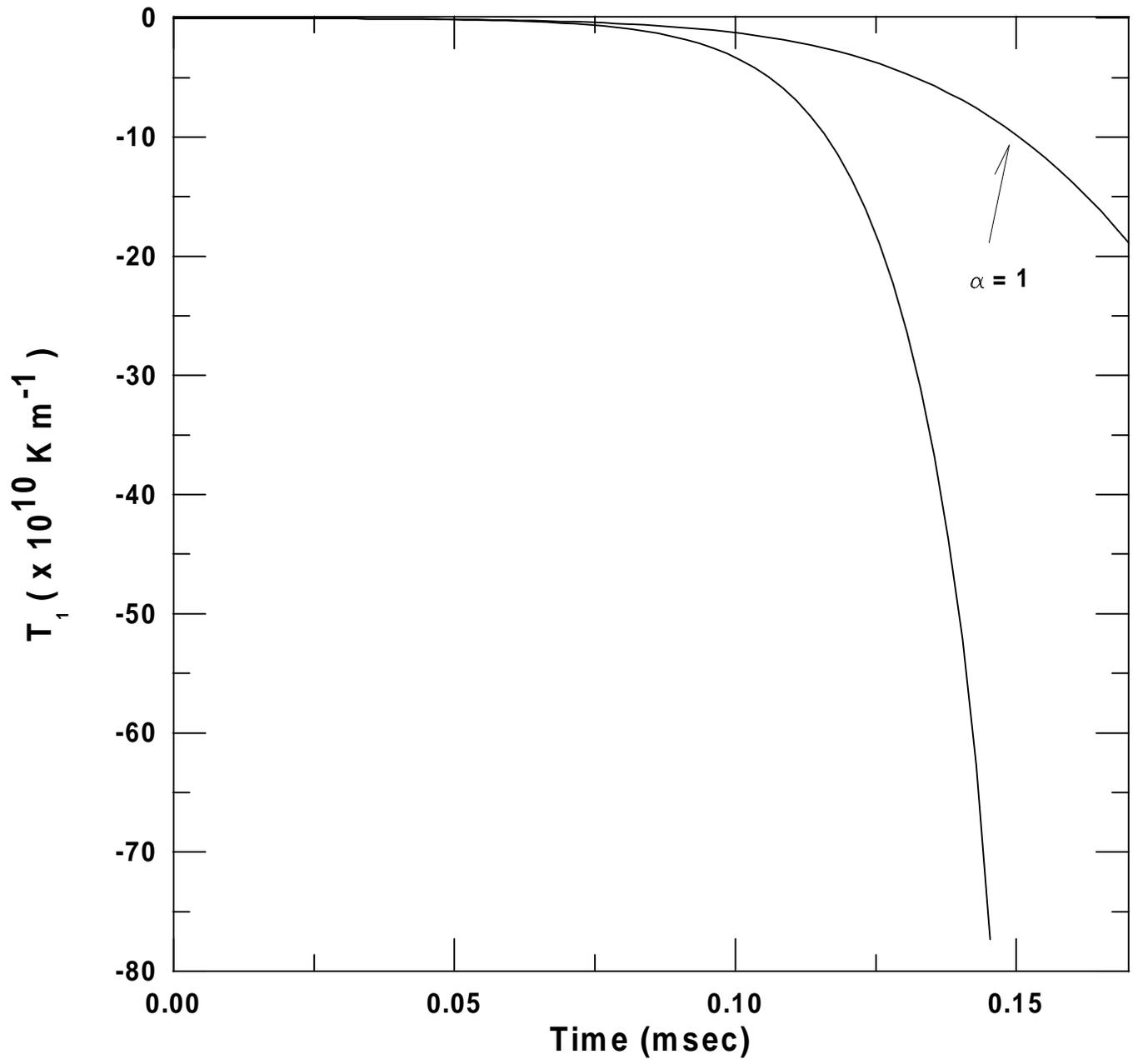
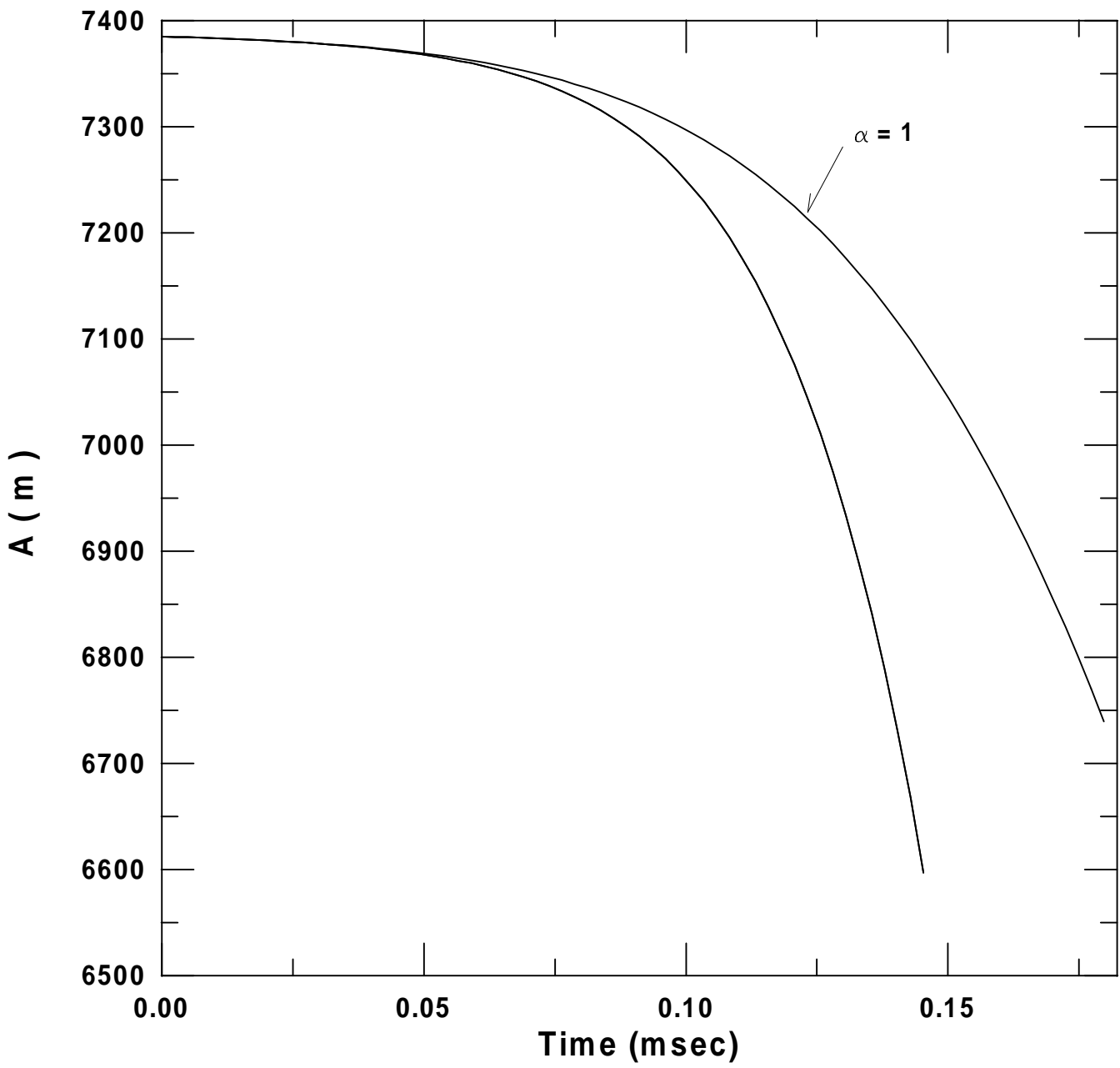
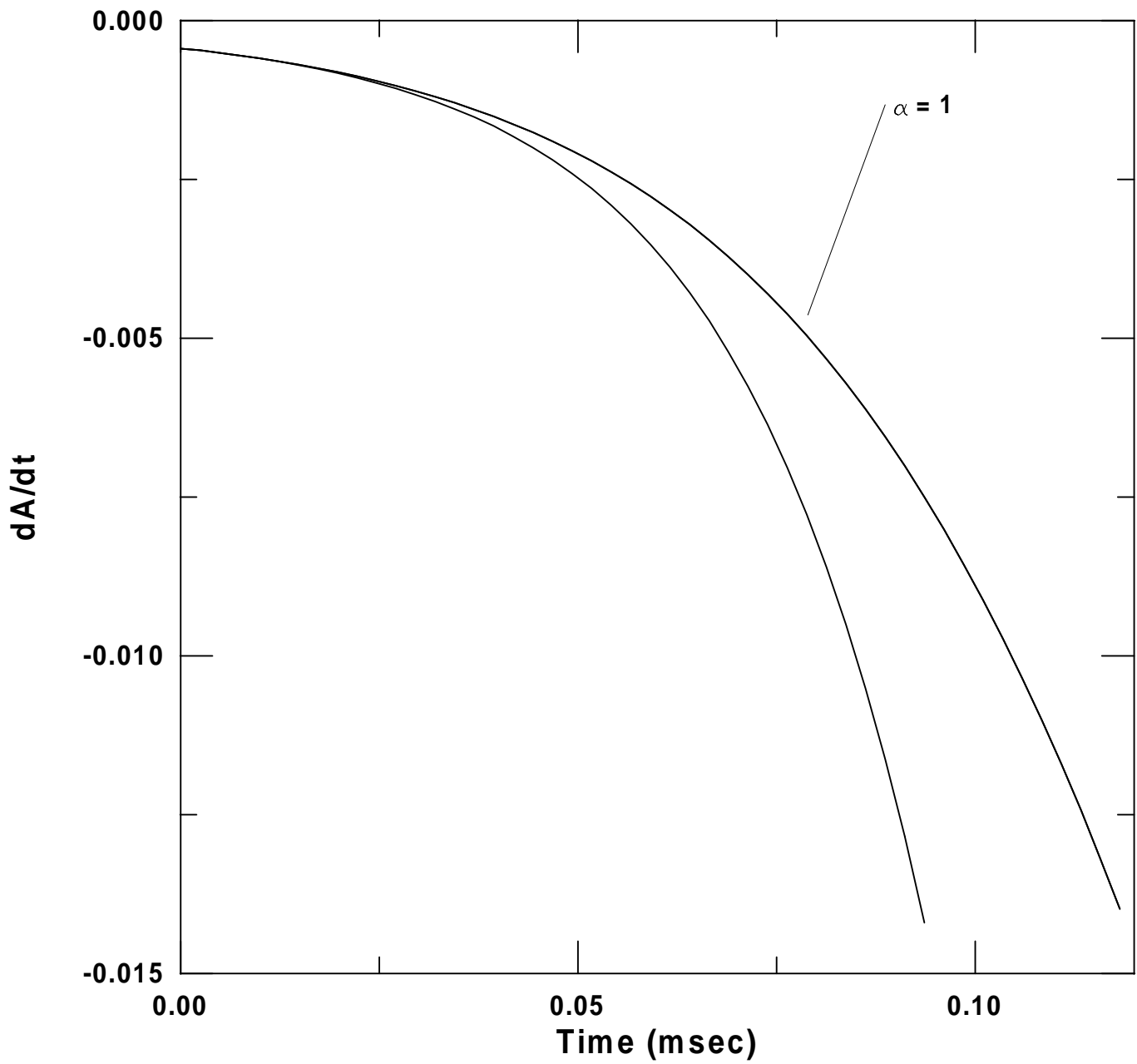
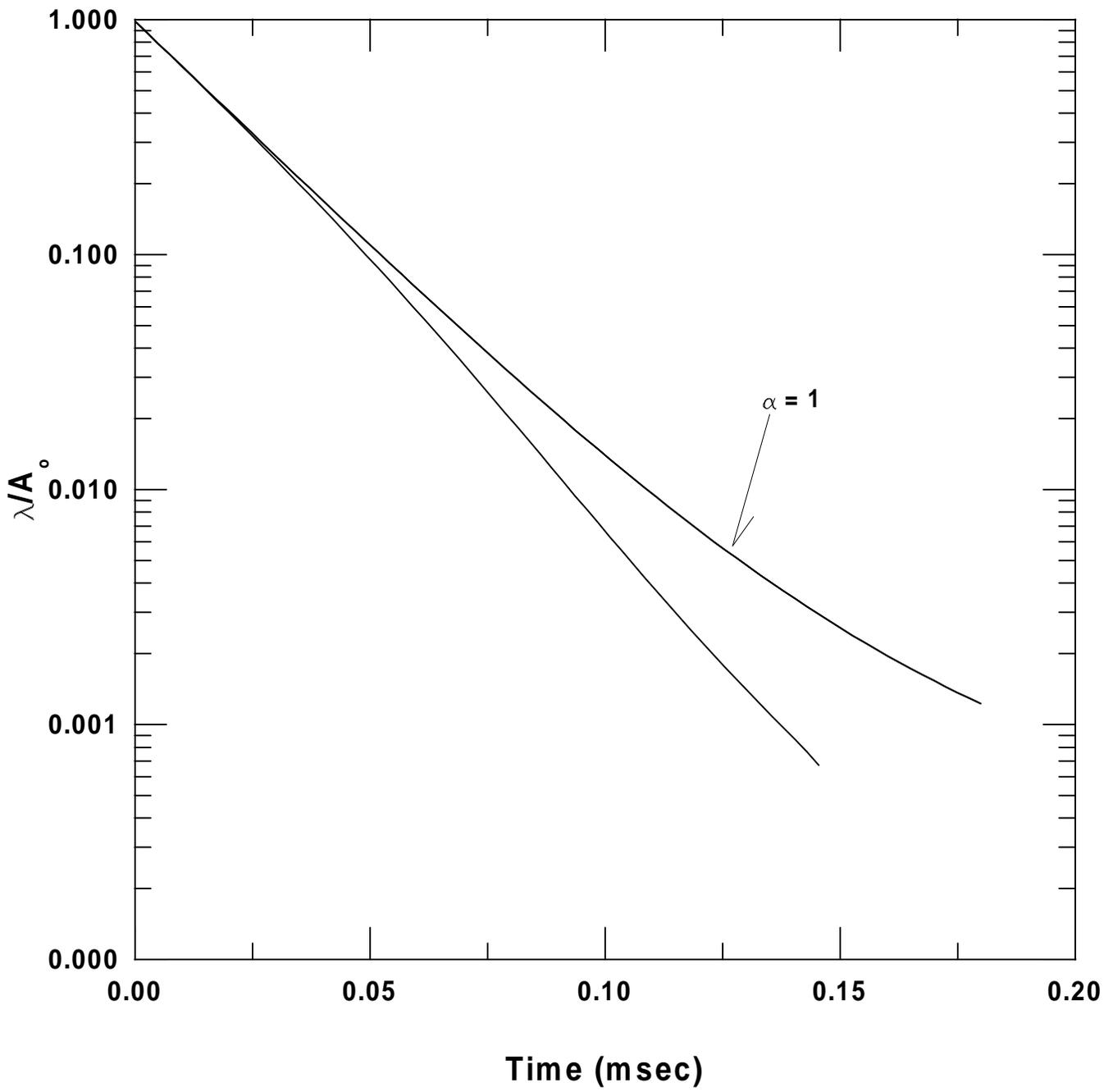
\end{document}